\documentclass[apjl]{emulateapj}

\slugcomment{}

\shorttitle{}
\shortauthors{Minezaki and Matsushita}

\begin{document}

\title{Comment on
``The dust sublimation radius as an outer envelope to the bulk of the narrow Fe K$\alpha$ line emission in Type 1 AGN''}

\author
{Takeo Minezaki\altaffilmark{1}
and Kyoko Matsushita\altaffilmark{2}
}

\altaffiltext{1}{Institute of Astronomy, School of Science, the University
of Tokyo, 2-21-1 Osawa, Mitaka, Tokyo 181-0015, Japan; minezaki@ioa.s.u-tokyo.ac.jp}
\altaffiltext{2}{Department of Physics, Tokyo University of Science,
1-3 Kagurazaka, Shinjuku-ku, Tokyo 162-8601, Japan}

\begin{abstract}
Recently, 
Gandhi, H\"onig, and Kishimoto submitted a manuscript to the arXiv
e-print service on the location of the emitting region of
the narrow FeK$\alpha $ line that appears
in the X-ray spectra of active galactic nuclei (AGNs)
compared with the inner radius of the dust torus (arXiv:1502.02661).
Prior to their manuscript, a similar discussion had already been presented
in a section of Minezaki \& Matsushita (2015),
which had been accepted for publication in the Astrophysical Journal.
Because Gandhi et al. made no reference to Minezaki \& Matsushita (2015)
apart from improperly citing it merely as an application of
the dust reverberation of AGNs,
we present a brief comparison of both papers.
Gandhi et al. compared the location of the FeK$\alpha$ emitting region
with the individually measured radius of the dust torus
for type 1 AGNs,
whereas Minezaki \& Matsushita (2015) examined it 
based on the scaling relation of the dust reverberation radius
for both type 1 and type 2 AGNs.
Nevertheless, Gandhi et al's main result is basically consistent with
and supports the results of Minezaki \& Matsushita (2015).
\end{abstract}


\section{Introduction}

Minezaki \& Matsushita (2015; hereafter MM15) proposed a new method
for estimating the mass of a supermassive black hole 
using the narrow core of the neutral FeK$\alpha $ emission line
in the X-ray spectra of active galactic nuclei (AGNs).
This method is applicable to obscured AGNs.
To construct their method,
MM15 first examined the location of the FeK$\alpha $ line emitting region,
which was estimated from the velocity width
of the neutral FeK$\alpha $ line core,
assuming the virial relation.
They compared the location of the FeK$\alpha $ line emitting region
with the reverberation radius of the optical broad emission line
and that of the near-infrared thermal emission from the inner dust torus.
MM15 concluded that the major fraction of
the neutral FeK$\alpha $ line core originates
between the outer broad emission-line region (BLR)
and the inner dust torus.

Immediately after MM15's manuscript appeared
in the arXiv e-print service as arXiv:1501.07522,
Gandhi, H\"onig, and Kishimoto submitted a manuscript to
the Astrophysical Journal Letters and to the arXiv e-print service
on the location of the emitting region of
the neutral FeK$\alpha $ line core (arXiv:1502.02661; hereafter GHK15).
They also estimated the location of the FeK$\alpha $ line emitting region
from the velocity width of the neutral FeK$\alpha $ line core
assuming the virial relation
and compared it with the radii of the BLR and the inner dust torus.
They concluded that the dust sublimation radius
forms an outer envelope to the bulk of the FeK$\alpha $ emission.

However, GHK15 cited MM15
only by a single sentence in the introduction section as follows: 
``{\it IR reverberation mapping is a growing field and provides
an alternate means to infer dust sizes and even constrain
black hole masses (e.g., Suganuma et al. 2006; Koshida et al. 2014;
Minezaki \& Matsushita 2015)}.''
Therefore, we briefly compare the GHK15 and MM15 manuscripts
to illustrate their similarities and differences.

\section{Discussion}
\subsection{Data and Targets}
MM15 and GHK15 adopted the FWHM data of the FeK$\alpha $ line core
acquired by the {\it Chandra} High Energy Transmission Grating Spectrometer
\citep{1994SPIE.2280..168M}, which affords the best
spectral resolution currently available in the FeK$\alpha $ energy band.

MM15 targeted seven type 1 AGNs and seven type 2 AGNs
whose black hole masses were available.
The FWHMs of the FeK$\alpha $ line core
were taken from \citet{2010ApJS..187..581S,2011ApJ...738..147S}.
For the type 1 AGNs, they selected only the targets
for which the best FWHM constraint for the neutral FeK$\alpha $ line core
was obtained by \citet{2010ApJS..187..581S}.

GHK15 targeted 13 type 1 AGNs
whose black hole masses and inner radii of the dust torus were available.
The inner radius of the dust torus was obtained
either by the dust reverberation or the near-infrared interferometry.
Type 2 AGNs were not included.

GHK15 obtained the value of the FWHM of the FeK$\alpha $ line core
mostly from \citet{2010ApJS..187..581S}.
Consequently, for six of their targets,
the FWHM data of the FeK$\alpha $ line core
matched those of MM15.
Because the best FWHM constraint was not available
for another five targets by \citet{2010ApJS..187..581S},
the FWHM data for them had relatively large uncertainties.
To obtain the upper limits on the FWHM 
for the remaining two targets,
GHK15 reanalyzed the data of \citet{2010ApJS..187..581S}
as well as the newly available data
\citep{2011MNRAS.414.1965L,2012ApJ...744L..21S}.

\subsection{Estimation of the FeK$\alpha $ Line Emitting Region}
To examine the location of the FeK$\alpha $ line emitting region
relative to the BLR,
MM15 compared the FWHMs of the neutral FeK$\alpha $ line core
and the broad Balmer emission lines
assuming the virial relation,
as performed in previous studies
\citep[e.g.,][]{2004ApJ...604...63Y,2006MNRAS.368L..62N,2010ApJS..187..581S,2011ApJ...738..147S}

Also assuming the virial relation,
GHK15 calculated the distance from the central engine
of the FeK$\alpha $ line emitting region as
$R_{\rm Fe}=(1/f')\times GM_{\rm BH}/({\rm FWHM}_{{\rm FeK}\alpha })^2 $,
where the factor $f'$ is determined from the cloud kinematics
in the the FeK$\alpha $ line emitting region.
They assumed $f'=3/4$ in the $R_{\rm Fe}$ calculation.
Next, they compared $R_{\rm Fe}$ with the BLR location,
$R_{\rm BLR}$, obtained from the reverberation radius
of the broad H$\beta$ emission line.

For almost all of GHK15's targets,
the black hole mass in the $R_{\rm Fe}$ calculation
was obtained from the reverberation of
the broad H$\beta$ emission line as
$M_{\rm BH}=f\times R_{\rm BLR}({\sigma }_{{\rm H}\beta} )^2/G$,
where $\sigma _{{\rm H}\beta}$ is the velocity dispersion
of the broad H$\beta$ emission line
and $f$ is the virial factor.
Because the FWHM of the broad H$\beta$ emission line
is basically proportional to its velocity dispersion, $\sigma _{{\rm H}\beta}$, 
the comparison of $R_{\rm Fe}$ and $R_{\rm BLR}$ in GHK15
can be equated with
that of the FWHMs of the neutral FeK$\alpha $ line core
and the broad H$\beta$ emission line,
as performed by MM15 and previous studies.

We note that MM15 and GHK15 assumed slightly different kinematics
of the FeK$\alpha $ emitting region and BLR.
If the FWHMs of the emission lines are compared in order to locate
the FeK$\alpha $ emitting region relative to the BLR,
the FeK$\alpha $ emitting region and BLR are 
implicitly assumed to have the same kinematics.
On the other hand,
if $R_{\rm Fe}$ is calculated from $M_{\rm BH}$,
the kinematics of the FeK$\alpha $ emitting region
are assumed as $f'=3/4$.

\subsection{Comparison with the Inner Radius of the Dust Torus}
MM15 examined the location of the FeK$\alpha $ emitting region
relative to the dust reverberation radius in the velocity domain.
To estimate the FWHM at the dust reverberation radius,
they scaled the FWHM of the broad H$\beta $ emission line
according to the virial relation.
The calculation was based on the systematic difference
between the reverberation radii of the broad H$\beta $ emission line
and the near-infrared dust emission,
which was derived by fitting the reverberation data of many AGNs
\citep{kosh+14}.

On the other hand, GHK15 compared $R_{\rm Fe}$ with
the inner radius of the dust torus ($R_{\rm dust}$),
which was measured by the dust reverberation \citep{kosh+14}
or the near-infrared interferometry
\citep{2009A&A...507L..57K,2011A&A...527A.121K,2013ApJ...775L..36K}.
These two measures of $R_{\rm dust}$
are known to systematically differ by about a factor of two
\citep[e.g.,][]{2009A&A...507L..57K,2011A&A...534A.121H,kosh+14,2014Natur.515..528H},
but GHK15 collected both types of data
as the dust sublimation radius.

The method of comparing the inner radius of the dust torus
with the FeK$\alpha $ emitting region is a major point of difference
between the MM15 and GHK15 papers;
MM15 based the comparison on
the scaling relation of the dust reverberation radius,
whereas GHK15 based it on individually measured values.

\subsection{Results}
For almost all their target AGNs,
MM15 found that the FWHM of the neutral FeK$\alpha $ line core
falls between the FWHM of the broad Balmer emission lines
and its corresponding value at the dust reverberation radius,
indicating that the major fraction of the neutral FeK$\alpha $ line core
originates between the outer BLR and the inner dust torus.
GHK15 found that $R_{\rm Fe}$ is never much larger than $R_{\rm dust}$,
indicating that the dust sublimation radius forms an outer envelope
to the bulk of the FeK$\alpha $ emission.
This is the key result of GHK15,
who compared $R_{\rm Fe}$ with $R_{\rm dust}$.

Clearly, the results of MM15 and GHK15 are consistent
at the outer boundary of the FeK$\alpha $ line emitting region.
For MM15's sole outlier, NGC 7469,
the FWHM of the neutral FeK$\alpha $ line core
exceeded that of the broad Balmer emission line.
Among MM15's targets, the FWHM of the neutral FeK$\alpha $ line core
was never much smaller 
than its corresponding value at the dust reverberation radius.

The results of MM15 and GHK15 also appear consistent at the inner boundary
of the FeK$\alpha $ line emitting region,
at least for the AGNs with the best FWHM constraints reported
by \citet{2010ApJS..187..581S}.
GHK15 argued that in these AGNs,
$R_{\rm Fe}$ matches $R_{\rm dust}$ well.
In fact, 
among the targets with the best FWHM constraints,
the $R_{\rm Fe}$ of NGC 3516 was smaller than $R_{\rm dust}$
but larger than $R_{\rm BLR}$,
while the $R_{\rm Fe}$ of NGC 7469 was much smaller than $R_{\rm dust}$
and was even smaller than $R_{\rm BLR}$.
As noted, NGC 7469,
whose FeK$\alpha $ line emitting region was determined as smaller than the BLR,
was the sole exception in MM15's target AGNs.

The situation is somewhat unclear for the other targets in GHK15,
because of the large uncertainties in the FWHMs of
their neutral FeK$\alpha $ lines.
Their $R_{\rm Fe}$ data were significantly scattered
from $R_{\rm dust}$ to much smaller values.
GHK15 suggested that the large uncertainties in the FWHM data
are not simply caused by low signal-to-noise ratios in the X-ray spectra
but reflect multiple source regions of the neutral FeK$\alpha $ lines.
As indicated in both studies,
substantial progress is expected in the near future
by the ASTRO-H X-ray satellite \citep{2010SPIE.7732E..0ZT},
which is capable of unprecedented energy-resolution spectroscopy
with superior sensitivity.

\end{document}